\documentclass{Interspeech2024}

\usepackage{amsmath,graphicx}
\usepackage{latexsym}
\usepackage{amsmath}
\usepackage{amssymb}
\usepackage{subfig}
\usepackage{float}
\usepackage{multirow}
\usepackage{pifont}
\usepackage{diagbox}

\interspeechcameraready

\title{Getting More for Less: Using Weak Labels and AV-Mixup for Robust Audio-Visual Speaker Verification}

\name{Anith}{Selvakumar}
\name{Homa}{Fashandi}

\address{
  LG Electronics, Toronto AI Lab, Canada}
\email{a.selvakumarasingam@lge.com, homa.fashandi@lge.com}


\begin{document}

\maketitle

\begin{abstract}
Distance Metric Learning (DML) has typically dominated the audio-visual speaker verification problem space, owing to strong performance in new and unseen classes. In our work, we explored multitask learning techniques to further enhance DML, and show that an auxiliary task with even weak labels can increase the quality of the learned speaker representation without increasing model complexity during inference. We also extend the Generalized End-to-End Loss (GE2E) to multimodal inputs and demonstrate that it can achieve competitive performance in an audio-visual space. Finally, we introduce AV-Mixup, a multimodal augmentation technique during training time that has shown to reduce speaker overfit. Our network achieves state of the art performance for speaker verification, reporting \textbf{0.244\%}, \textbf{0.252\%}, \textbf{0.441\%} Equal Error Rate (EER) on the VoxCeleb1-O/E/H test sets, which is to our knowledge, the best published results on VoxCeleb1-E and VoxCeleb1-H.
\end{abstract}

\section{Introduction}

Human interactivity with artificially intelligent systems continue to gain popularity, especially as devices become capable of advanced tasks and are seamlessly integrated with our daily lives (e.g., digital assistants). A critical component to enabling personalized interactions with such systems is speaker verification, the process of identifying whether a speaker matches with a pre-enrolled speaker profile. Applications can range from user authentication to personalized experiences, however, environmental noise and visual occlusion are only a few of the challenges associated with performing reliable speaker verification in real-world settings \cite{Shon2019}. To this end, multimodal systems have been increasing in popularity due to the potential of added robustness and improved performance \cite{Huang2021}. For the case of speaker verification, leveraging both audio and visual modalities, in particular, have shown improvement in false-reject rates when compared to audio-only and visual-only systems \cite{Shi2022} \cite{Qian2021} \cite{Sari2021} \cite{Tao2020} \cite{Chen2020}.

This work explores to improve upon the existing benchmarks in audio-visual speaker verification without expending on model complexity, by introducing data-efficient training and augmentation strategies. Our main contributions are as follows:

\begin{enumerate}
  \item We demonstrate that multi-task learning with inexpensively-obtained, weak labels can be used to enhance the representations learned by DML.
  \item We extend the Generalized End to End Loss (GE2E) \cite{Wan2018} from a unimodal to a multimodal input space, and for the first time, validate its efficacy beyond an audio-only task.
  \item We introduce AV-Mixup, a multimodal augmentation strategy to reduce speaker overfit and improve generalization.
\end{enumerate}

The collection of these contributions yield SOTA performance with EER of 0.244\%, 0.252\%, and 0.441\% on the VoxCeleb1-O/E/H test splits.

\section{Background}

\subsection{Related Works}

Prior studies reveal interesting developments in audio-visual speaker verification. However, a majority rely on traditional DML training approaches, and in general, devise complex networks or rely on expensive data collection to realize modest performance gains. For example, Qian et al. \cite{Qian2021} introduced a joint learned embedding-level network architecture, trained with their contrastive loss sampling and data augmentation strategy originally presented in \cite{Chen2020}. Sun et al. \cite{Sun2022} implemented joint-attention pooling on the audio-visual inputs that enhance the weights of impactful time frames. Tao et al. \cite{Tao2022} cited noisy labels in large-scale datasets as a significant limitation, and proposed a two-step multimodal deep cleansing network to identify and remove noisy training samples. Finally, Lin et al. \cite{Lin2023} introduced a large-scale dataset that showed improved performance when used as a supplementary training set, with best results on their ResNet with frequency-wise Squeeze-Excitation model (denoted M3).

Our work differs by hypothesizing that DML can be enhanced by introducing even a weakly supervised multi-task component to the objective function. This is on the basis of \cite{Kobs2021}, where it was shown that the features learned through classification and contrastive approaches can differ. Further, rather than extensive data collection or dataset cleansing, we hypothesize that noisy labels can be beneficial to achieve more robust speaker representations, by extending training methods and proposing novel augmentation techniques that collectively serve as regularization during training for open-set tasks.


\subsection{Multimodal Fusion}

\begin{figure*}[t!]
\centerline{\includegraphics[height=1.30 in]{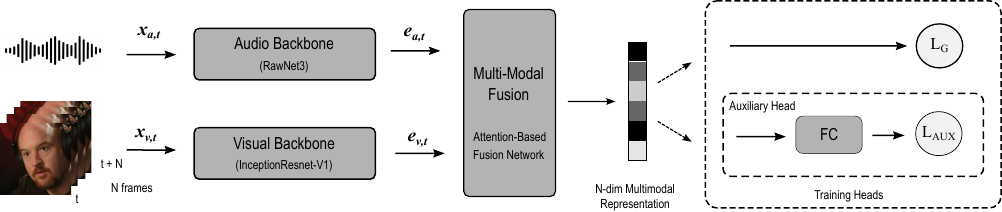}}
\caption{System level diagram of the REPTAR network. The multimodal representation feeds into the training heads, which are removed during inference time. For inference, the multimodal representation can be used directly for speaker verification} \label{arch}
\end{figure*}

Multimodal fusion has been achieved through many different techniques \cite{Shon2019} \cite{Fukui2016} \cite{Arevalo2019}. For robustness in non-ideal scenarios in the audio-visual domain, the attention-based fusion network (AFN) proposed in \cite{Shon2019} is of particular interest. AFN learns to adapt to corrupt or missing modalities by re-weighing the contribution of either modality at time of fusion. This is through an attention mechanism that extended across the modality axis to obtain modality attention weights:

\begin{equation}
\hat{a}_{\{a,v\}}=f_{att}([\textbf{e}_a,\textbf{e}_v]) = \textbf{W}^T[\textbf{e}_a,\textbf{e}_v] + \textbf{b},
\label{attn_eqn}
\end{equation} 

where ${e_a}$ and ${e_v}$ are transformed audio and visual representations, respectively. Learnable parameters $\textbf{W}^T$ and $\textbf{b}$ are optimized during the training process. Modality attention weights are then re-scaled via Softmax to obtain scores between [0, 1] and applied across the embedding axis prior to aggregation to form the fused multimodal representation.


\subsection{Generalized End-to-End (GE2E) Loss}

The GE2E loss, originally proposed in \cite{Wan2018} for audio-based speaker verification, adopts an approach that uses class centroid distances during optimization. Specifically, the loss is calculated from a similarity matrix between each utterance representation embedding to all speaker utterance centroids. From this matrix, a total contrastive loss is calculated based upon positive components and a hard negative component.

\section{Robust Audio-Visual Person Encoder}

\subsection{Generalized End-to End Multimodal Loss}

Large scale datasets often contain noisy labels that can confuse networks during training and limit performance. We, however, propose to leverage these noisy samples to improve generalizability. We hypothesize that using a centroid-based optimization approach, outliers and noisy labels will act as a regularizer during training to lead to better generalization. We achieve this through extending the GE2E loss to a multimodal input space, and refer to this new loss function as GE2E-MM.

The GE2E-MM architecture relies on batching $N \times M$ audio and visual inputs, $x_{\{a,v\},ji}$ $(1 \le j \le N, 1 \le i \le M)$, where $N$ and $M$ are unique speakers and speaker utterances, respectively. 

We define the audio-visual latent representation as:

\begin{equation}
\textbf{e}_{ji} = \frac{  f(\textbf{x}_{a,ji} ; \textbf{x}_{v,ji} ; \textbf{w}) } {  || f(\textbf{x}_{a,ji} ; \textbf{x}_{v,ji} ; \textbf{w}) ||_2 }
\label{transerfunction}
\end{equation}
where $f(\textbf{x}_{a,ji} ; \textbf{x}_{v,ji} ; \textbf{w})$ represents the transfer function of the neural network, with $\textbf{x}_{a,ji}$ and $\textbf{x}_{v,ji}$ representing raw audio and visual inputs; and \textbf{w} representing the network weights. Using this, a similarity matrix, $S_{ji,k}$, of scaled cosine similarities is computed, representing a similarity metric between each multimodal embedding $e_{ji}$ and each speaker centroid, $c_k$, from the $N \times M$ batch: 

\begin{equation}
\textbf{c}_{k} = \frac{1}{M}\sum_{m=1}^{M} \textbf{e}_{km}
\label{centroid}
\end{equation}

\begin{equation}
\textbf{S}_{ji,k} = w \cdot cos(\textbf{e}_{ji}, \textbf{c}_k) + b\\
\label{similarity_eqn}
\end{equation}


where $w$, $b$ are learnable parameters. Using this similarity matrix, a contrastive loss is calculated for each multimodal representation, $e_{ji}$, focusing primarily on all positive pairs and a hard negative pair. The GE2E-MM loss, $\mathcal{L}_{G}$, is then defined as:

\begin{equation}
\mathcal{L}_{G}(\textbf{S}) = \sum_{j,i} \mathcal{L}(\textbf{e}_{ji})
\label{embedding_loss_0}
\end{equation}
where,

\begin{equation}
\mathcal{L}(e_{ji}) = 1 - \sigma(\textbf{S}_{ji,j}) + \max_{\substack {1\le k\le N\\ k \neq j}} \sigma (\textbf{S}_{ji,k})
\label{embedding_loss_1}
\end{equation}

where $\sigma$ represents the sigmoid function. Optimization of Equation \ref{embedding_loss_1} has the effect of pushing embeddings from identical speakers towards its respective centroid, and away from its closest dissimilar speaker centroid. 

\subsection{Multi-Task Objective Function}

We hypothesize that adding an age classification task will force more subtle characteristics to be extracted from the inputs and embedded in the multimodal representation. With this auxiliary task, we can define a multi-task loss function:

\begin{equation}
\mathcal{L}_{MTL} = \gamma \cdot \mathcal{L}_{G}(S) + (1 - \gamma) \cdot \mathcal{L}_{AUX},
\label{objective_function}
\end{equation}
\\
where $\gamma$ is a scalar weight that is applied in order to balance the losses and prevent one task from dominating. The parameter is obtained through hyper-parameter tuning. $\mathcal{L}_{G}$ and $\mathcal{L}_{AUX}$ are GE2E-MM and auxiliary task losses, respectively. 

\subsection{AV-Mixup for Multimodal Augmentation}

Audio and visual samples sourced from the same utterance can be highly correlated on speaker-irrelevant features \cite{mun2022} \cite{Nagrani2020}. This can limit the learning of distinctive features and instead cause the training to focus on peripheral attributes such as noise or environmental factors. To prevent this, we propose AV-Mixup, an augmentation technique whereby unique audio-visual pairs are recreated using audio and visual samples that are extracted from disjoint utterances from the same speaker.

Using the AFN for audio-visual fusion, we implemented the above-mentioned developments to form the end-to-end encoder network: `\textbf{R}obust \textbf{E}ncoder for \textbf{P}ersons through Learned Multi-\textbf{TA}sk \textbf{R}epresentations' (REPTAR), shown in Figure \ref{arch}.

\section{Experimentation}

\subsection{Setup}

\subsubsection{Dataset and Preprocessing}

VoxCeleb is a large-scale audio-visual dataset for speaker recognition and is widely used in literature. For our experimentation, VoxCeleb2 \cite{Chung2018} was used for training while pre-defined test splits from VoxCeleb1 \cite{Nagraniy2017} were used for evaluation. 

During pre-processing, utterance video files were decomposed into face tracks at a 1 frame per second (FPS) rate and re-scaled to $160 \times 160$ pixels, and a voice track cropped to a random window size between 4-8 seconds. For evaluation, utterances were sourced from the predefined test splits. Voice clips were limited to a fixed 4 second window with a random onset time and faces were extracted from the same utterance.

\subsubsection{Voice and Face Representation}

Pre-trained encoders were used to obtain voice and face representations. For voice profiles, RawNet3, proposed in \cite{Jung2022} was used. RawNet3 was trained on the VoxCeleb2 dataset and evaluated on the VoxCeleb1-O test split and demonstrates competitive EER performance. For face representation, the InceptionResnet-V1 architecture \cite{Szegedy2017} was used, which was pre-trained on the VGGFace2 \cite{Cao2018} dataset. VGGFace2 was confirmed to have a negligible overlap between the test set (5 out of 1251 speakers), which was verified to play no impact to the model performance when rounded to three significant figures.

\subsubsection{Multimodal Fusion}

For our implementation of the AFN, the 512 dimension face embedding and 256 dimension voice embedding obtained from their pre-trained models are L2 normalized and transformed independently into a 512 dimension space for equal representation prior to fusion. This transformation consists of two linear layers, with a ReLU activation and batch normalization layer in between. The attention layer is implemented as a linear layer with input size 1024 and output of 2 to represent the modality scores. These softmaxed scores are applied as multiplicative factors to the transformed representations, which are then concatenated to form the multimodal representation.

\subsubsection{Weakly-Supervised Auxiliary Task}

An age prediction auxiliary task was implemented to enhance feature learning during training. The task head consisted of two linear layers with a ReLU activation and batch normalization layer in between. A sigmoid layer was used at the end of the network to represent normalized age predictions. Mean-squared error loss was used as the objective function. The optimal $\gamma$ value in the compound loss function of Equation \ref{objective_function} was determined to be 0.015 through hyperparameter tuning.

Age labels were obtained from the AgeVoxCeleb dataset \cite{Tawara2021}, which contain estimated ages for approximately 5000 of 6112 speakers of the VoxCeleb2 dataset. These labels can be considered weak due to label inaccuracies and incompleteness. 

\subsubsection{Training and Evaluation}

Our proposed REPTAR model was trained on a Tesla V100 GPU. Batch size was set to 64 with 10 utterances per speaker. The Adam optimizer \cite{Kingma2015} was used with an exponentially decaying learning rate, initially set to 0.05 and decaying at a factor of 0.9 per epoch. Early stopping with a patience of 5 epochs was used to prevent overfitting. Multiple seeds were tested to demonstrate reproducibility of results.

The Equal Error Rate (EER) metric was calculated for each of the VoxCeleb1 test splits to evaluate the performance of the speaker verification system. EER corresponds to the error rate at which the False Positive Rate (FPR) and False Negative Rate (FNR) is equal, and is a standard metric used for speaker verification \cite{Nagraniy2017}.


\begin{figure}[t]
\centering
\subfloat[\centering Triplet Loss]{{\includegraphics[width=1.3in]{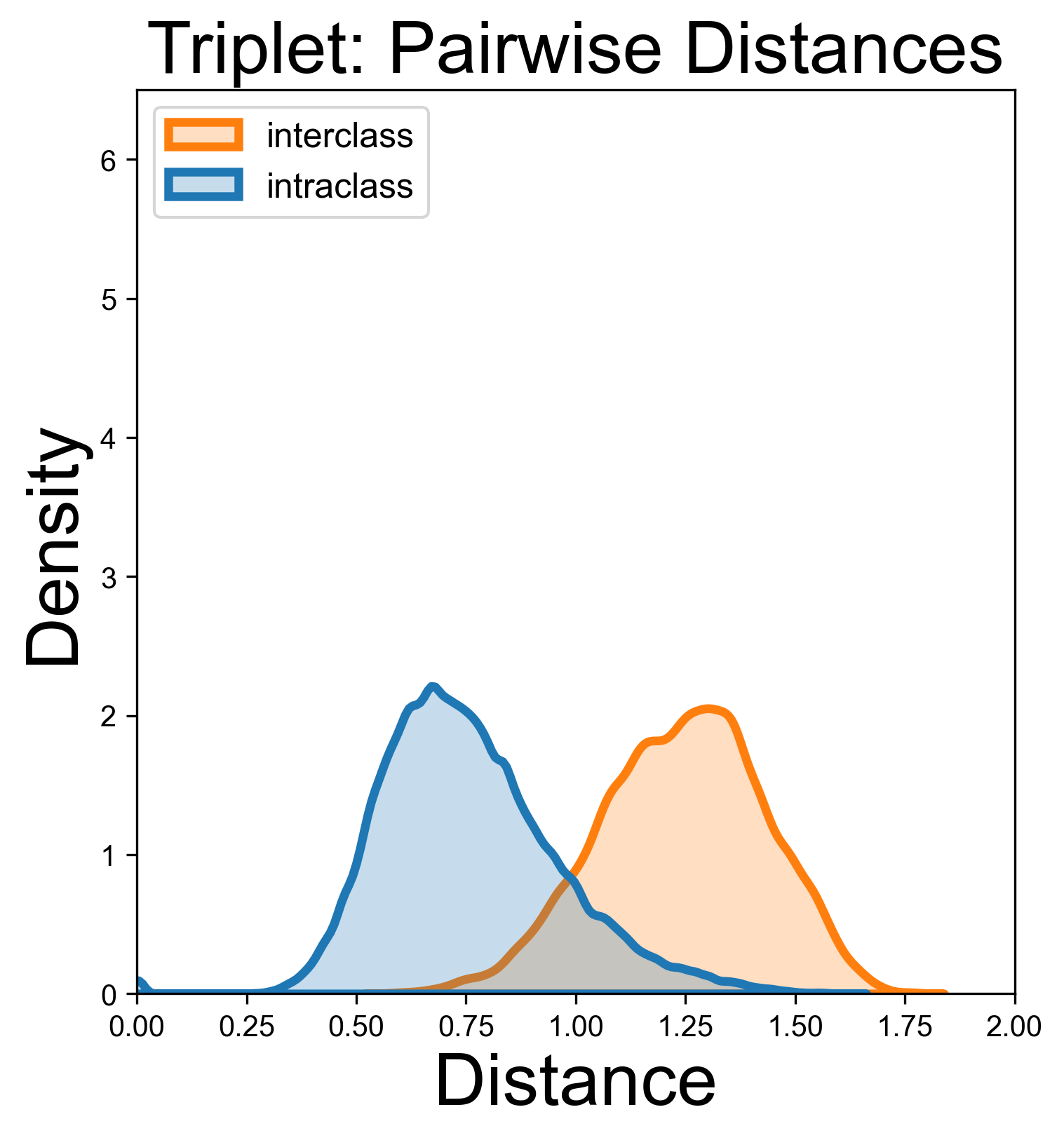} }}%
\qquad
\subfloat[\centering GE2E-MM Loss]{{\includegraphics[width=1.3in]{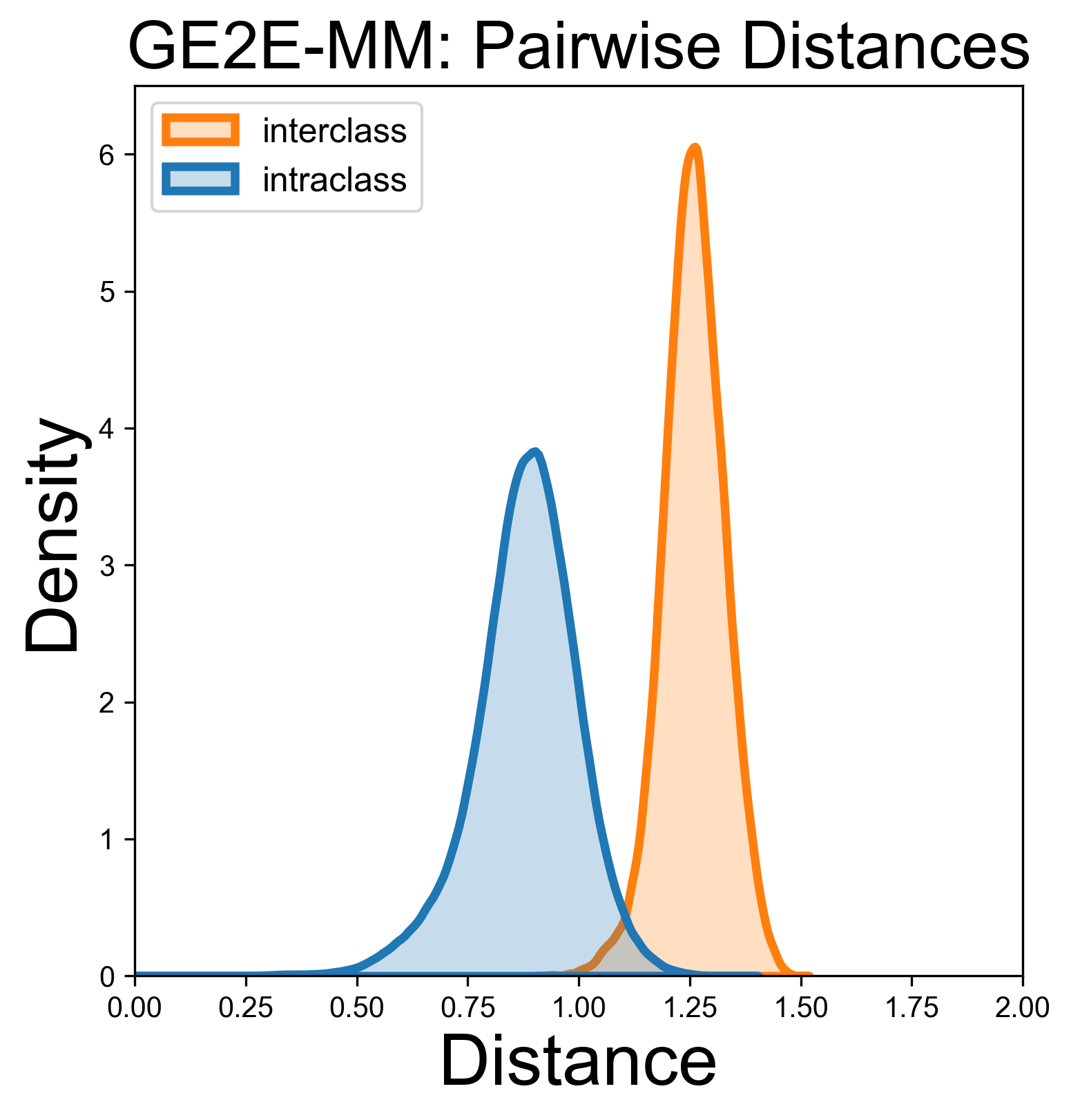} }}%
\caption{Interclass and Intraclass Pairwise Distance Distribution of Triplet and GE2E-MM models for a random speaker.}%
\label{ablation_distribution}%
\end{figure}


\begin{table}
\begin{center}
{\caption{Quality of encoded person representation clusters against triplet and GE2E-MM loss (with, without auxiliary task) }\label{cluster_table}}
\begin{tabular}{ c c c c}
\hline
\rule{0pt}{12pt}
\multirow{2}{*}{Clustering Metric} & \multirow{2}{*}{$\mathcal{L}_{triplet}$} & \multicolumn{2}{c}{$\mathcal{L}_{GE2E-MM}$} \\ 
& & $_{No Aux}$ & $_{Aux}$ \\
\hline
\\[-6pt]
Silhouette Score \cite{Rousseeuw1987} ($\uparrow$) & 0.233 & 0.501 & 0.504\\ 
Calinski-Harabasz \cite{Calinski1974} ($\uparrow$) & 4541 & 6264 & 6454\\
Davies-Bouldin \cite{Davies1979} ($\downarrow$) & 1.4657 & 0.8613 & 0.8599\\
\hline
\hline
\end{tabular}
\end{center}
\end{table}

\subsection{Results and Ablation Studies}

We performed extensive experiments to study the effect of each of the proposed strategies. This section describes the results obtained along with a comparison to other relevant works.

\begin{table}
\begin{center}
{\caption{Ablation study showing the effect of proposed loss function, auxiliary task training, and AV-Mixup on EER.}\label{table_overview}}
\begin{tabular}{ c c c c c c}
\hline
\rule{0pt}{12pt}
\multirow{2}{*}{Loss} & \multicolumn{2}{c}{Training Config} & \multicolumn{3}{c}{Evaluation (VC1)} \\
\cline{2-3}
\cline{4-6}
\\[-6pt]
& Aux & AV-Mix & O & E & H \\

\hline
Triplet \cite{Schroff2015} & N & Y & 6.85 & 9.57 & 5.16 \\  
GE2E-MM & N & N & 0.427 & 0.321 & 0.568 \\  
GE2E-MM & N & Y & 0.323 & 0.292 & 0.507 \\  
\textbf{GE2E-MM} & \textbf{Y} & \textbf{Y} & \textbf{0.244} & \textbf{0.252} & \textbf{0.441} \\
\hline
\hline
\end{tabular}
\end{center}

\begin{center}
{\caption{EER of REPTAR in the presence of corrupted and missing modalities on the VoxCeleb1-O test split}\label{partial_input_results}}
\begin{tabular}{c c c | c}
\hline
\rule{0pt}{12pt}
Architecture & Audio Input & Visual Input & \textbf{EER} \\
\hline
& \textit{clean} & \textit{clean} & 0.24 \\
& \textit{corrupt} & \textit{clean} & 1.98 \\
REPTAR & \textit{missing} & \textit{clean} & 1.12 \\
& \textit{clean} & \textit{corrupt} & 6.12 \\
& \textit{clean} & \textit{missing} & 1.64 \\
\hline
\hline
\end{tabular}
\end{center}
\end{table}

\subsubsection{Effect of the GE2E-MM Loss:}

The impact of the GE2E-MM loss was analyzed by studying the encoded audio-visual representations of speaker utterances from identical networks that differed only on the objective function they were trained on (GE2E-MM or Triplet \cite{Schroff2015}). Intraclass pairwise Euclidean distance was measured between all combinations in the set of utterance encodings obtained from the same speaker. Similarly, interclass pairwise distance was measured as the Euclidean distance between the reference speaker's encoded utterance \textit{centroid} with that of the utterance centroid for every other speaker in the set. 

The intraclass and interclass sample distributions for the triplet loss and GE2E-MM loss models are shown in Figure \ref{ablation_distribution}. Results show significant reduction in the overlap between intraclass and interclass distributions on the GE2E-MM loss trained network, implying more compact person representations. To validate this further, we employ the Silhouette coefficient \cite{Rousseeuw1987}, Calinski-Harabasz score \cite{Calinski1974}, and the Davies-Bouldin score \cite{Davies1979} on the test set. As shown in Table \ref{cluster_table}, the GE2E-MM loss show improved cluster quality for all three metrics.

\subsubsection{Effect of the Weakly-Supervised Auxiliary Task and AV-Mixup Multimodal Augmentation:}

The results in Table \ref{cluster_table} show the improvement of cluster quality when comparing to the same network trained without the additional task loss. This sentiment is echoed in the results of Table \ref{table_overview}, showing an average 17\% EER improvement on the VC1-O/E/H test splits. We believe that the age classification auxiliary task ensured that distinctive markers from both modalities are preserved in the multimodal representation to help generalization and improve overall performance.

\begin{table}[t]
\begin{center}
{\caption{EER measurements on various training dataset configurations. Best results per evaluation split are in \textbf{bold} }\label{table_voxblink}}
\begin{tabular}{ c  c c  c c c}
\hline
\rule{0pt}{12pt}

\multirow{2}{*}{Model} & \multicolumn{2}{c}{Train Set} & \multicolumn{3}{c}{Evaluation (VC1)} \\
\cline{2-3}
\cline{4-6}
\\[-6pt]
& VC2 & VB & O & E & H \\

\hline
Lin et al (M3)\cite{Lin2023} & Y & N & 0.622 & 0.761 & 1.391 \\  
REPTAR & Y & N & 0.244 & \textbf{0.252} & \textbf{0.441} \\
\hline
Lin et al (M3)\cite{Lin2023} & Y & Y & 0.441 & 0.681 & 1.268 \\  
REPTAR & Y & Y & \textbf{0.196} & 0.316 & 0.537 \\
\hline
\hline
\end{tabular}
\end{center}
\end{table}




\begin{table}[ht]

\begin{center}
{\caption{Proposed model EER performance compared to SOTA. Lower value signifies a better result. Best results are in \textbf{bold}}\label{table_comparison}}

\begin{tabular}{c c c c c}
\hline
\rule{0pt}{12pt}

\rule{0pt}{12pt}
\multirow{2}{*}{Model} & \multirow{2}{*}{Modality} & \multicolumn{3}{c}{VoxCeleb1} \\
\cline{3-5}
\\[-6pt]
& & O & E & H \\

\hline
& A & 2.31 & 2.23 & 3.78 \\ 
Chen et al \cite{Chen2020}& V & 2.26 & 1.54 & 2.37 \\ 
& AV & 0.585 & 0.427 & 0.735 \\
\hline
& A & 1.62 & 1.75 & 3.16 \\ 
Qian et al \cite{Qian2021} & V & 3.04 & 2.18 & 3.23 \\ 
& AV & 0.558 & 0.441 & 0.793\\
\hline
& A & \textbf{0.99} & 1.24 & 2.27 \\ 
Sun et al \cite{Sun2022} & V & 1.44 & 1.28 & 2.14\\ 
& AV & \textbf{0.18} & 0.26 & 0.49\\
\hline
Lin et al (M3) \cite{Lin2023} & AV & 0.622 & 0.761 & 1.39\\
\hline
Lin et al (M4) \cite{Lin2023} & AV & 0.580 & 0.775 & 1.44\\
\hline
& A & 1.64 & \textbf{1.12} & \textbf{1.85} \\ 
REPTAR & V & \textbf{1.12} & \textbf{1.19} & \textbf{1.82} \\ 
& AV & 0.244 & \textbf{0.252} & \textbf{0.441} \\ 
\hline
\hline
\end{tabular}

\end{center}
\end{table}







Through randomization of visual and audio speaker inputs by AV-Mixup, we were able to see an average 15\% improvement in performance compared to a model trained using audio and visual inputs from the same utterance. Similar to the findings of Nagrani et al. \cite{Nagrani2020} on disentangled linguistic content and speaker identity yielding better generalization, we believe our improvement is also as result of minimizing mutual information, but within the speaker audio-visual input space.

\subsubsection{Effect of Corrupted and Missing Modality:}

Measuring robustness to non-ideal situations was performed by recreating absent or corrupt modalities. An absent modality was emulated by setting the input to zero. A corrupt input was emulated using additive white Gaussian noise (AWGN), with $\mu = 0$ and $\sigma_{v} = {[0,255]}$ or $\sigma_{a} = {[-1,1]},$ where $\mu$ and $\sigma$ are mean and standard deviation, respectively. This methodology is consistent with existing literary works \cite{Shon2019}.

The results in Table \ref{partial_input_results} show that the multimodal network is robust to missing or corrupt inputs without significantly compromising performance. This can be compared to other architectures that rely on both modalities to be present, a constraint that is not always feasible in real-world scenarios.

\subsubsection{Effect of Additional Training Data:}

The VoxBlink-clean (VB) dataset was explored as a potential complementary training dataset \cite{Lin2023}, and was compared to against the reported benchmarks. Results are shown in Table \ref{table_voxblink}. Despite the additional set of 1.45M utterances across 38K new speakers, a degradation of performance on the VC1-E/H splits was observed. This leads us to believe that the training and optimization strategies proposed (i.e. MTL, AV-Mixup) in REPTAR can be seen as an alternative, more data-efficient way to improve model performance.




\subsubsection{Summary of Results}

Our proposed model REPTAR achieves competitive performance against the previous state of the art, yielding best published results in 7 of 9 test configurations. Results are shown in Table \ref{table_comparison} along with a comparison to related works. The results of Tao et al. \cite{Tao2022} were omitted due to train-test contamination.

Our proposed model is able to achieve SOTA performance on all test configurations of the VoxCeleb1-E and VoxCeleb1-H test splits, which are considerably larger and targets a broader demographic compared to the VoxCeleb1-O split. Jointly, VoxCeleb1-E and VoxCeleb1-H can be used to describe the REPTAR’s generalizeability and quality of feature extraction. The results show that REPTAR goes beyond encoding basic high-level features such as nationality and gender in the representation space, and is able to extract features that can be used to distinguish between even the most similar of speakers.


\section{Conclusion}

In this paper we explored data-efficient approaches to improving the robustness of speaker verification systems. Specifically, we demonstrated how DML representation learning can be enhanced, by introducing an auxiliary task trained on inexpensive, weak labels and measuring the quality of the resulting speaker representations. We also show how noise in the training set can be leveraged to improve generalization by introducing the GE2E-MM loss as well as AV-Mixup, a multimodal data augmentation technique. A comprehensive study of MTL task selection and task weighting strategies is left as future work.

\bibliographystyle{IEEEtran}
\bibliography{mybib}

\end{document}